\documentclass[aps,showpacs,a4paper,floatfix,twocolumn,prc,amsmath,amssymb]{revtex4-1}
\usepackage{graphicx}
\usepackage{epsfig}
\usepackage{ulem}
\usepackage{morefloats}
\usepackage{rotating}
\usepackage{color}
\usepackage{soul}

\def\beq{\begin{equation}}
\def\eeq{\end{equation}}
\def\beqa{\begin{eqnarray}}
\def\eeqa{\end{eqnarray}}
\def\ban{\begin{eqnarray*}}
\def\ean{\end{eqnarray*}}
\def\bi{\begin{itemize}}
\def\ei{\end{itemize}}

\newcommand{\pc}[1]{\ensuremath{\left(#1\right)}}

\definecolor{mypink}{RGB}{219, 48, 222}

\begin{document}

\title{Light clusters in warm stellar matter: explicit mass shifts and universal cluster-meson couplings}
 \author{Helena Pais$^1$}
 \author{Francesca Gulminelli$^2$}
 \author{Constan\c ca Provid{\^e}ncia$^1$}
 \author{Gerd R\"opke$^{3,4}$}

 \affiliation{$^1$CFisUC, Department of Physics, University of Coimbra,
   3004-516 Coimbra, Portugal. \\
 $^2$LPC (CNRS/ENSICAEN/Universit\'e de Caen Normandie), UMR6534, 14050 Caen c\'edex, France. \\
 $^3$Institut f\"ur Physik, Universit\"at Rostock, D-18051 Rostock, Germany. \\
 $^4$National Research Nuclear University (MEPhI), 115409 Moscow, Russia.}

\begin{abstract}
In-medium modifications of light cluster properties in warm stellar matter are studied within the relativistic mean-field approximation. In-medium effects are included by introducing an explicit binding energy shift analytically calculated in the Thomas-Fermi approximation, supplemented with a phenomenological modification of the cluster couplings to the $\sigma$ meson. A linear dependence on the $\sigma$ meson is assumed for the cluster mass, and the associated coupling constant is fixed imposing that the virial limit at low density is recovered. The resulting cluster abundances come out to be in reasonable agreement with constraints at higher density coming from heavy ion collision data. Some comparisons with microscopic calculations are also shown. 
\end{abstract}

%\pacs{21.60.-n,21.60.Ev,26.60.Gj,24.10.Jv}

\maketitle

\section{Introduction}

Neutron stars are compact objects where a wide range of densities, pressures
and temperatures can be achieved. In the outer part of the star, where 
the baryonic density is below the central density of atomic nuclei, matter is inhomogeneous
and clusterized into nuclei. If matter is catalyzed as in equilibrium neutron stars, the crust is
a Wigner solid, and the nuclear components are sufficiently heavy to be treatable within the density functional theory \cite{DFT}. However, above the crystallization temperature, the crust melts and  
light clusters with only a few number of nucleons contribute to the equilibrium \cite{typel10}. At sufficiently low density and high temperature, corresponding to a large fraction of the post-bounce supernova dynamics, they  constitute the main baryonic contribution \cite{ccsn}, and we can expect that these light particles will play an important role in the neutron star cooling, accreting systems, and binary mergers.

The light clusters will eventually melt,  when high enough temperatures
are achieved, but in either warm neutron stars, where $T\lesssim 2$ MeV, or
core-collapse supernova environments, where $T\lesssim20$ MeV,  or
binary star mergers with $T\lesssim10$ MeV, they can appear, as these environments gather the perfect conditions for their formation. Moreover, all these clusters, light and heavy, may have a non-negligible effect in the core-collapse supernova mechanism \cite{ccsn}, as they affect the neutrino mean free path, and consequently, the cooling of the star. 

Light clusters in nuclear matter have been included within different
approaches: in the single nucleus approximation (SNA), such as in the Lattimer and Swesty
(LS) \cite{ls} equation of state (EoS) based on the compressible
liquid droplet model  or the Shen \cite{shen} EoS using a
relativistic mean-field (RMF) model,  non-homogeneous matter is
described in the Wigner Seitz approximation,  with a  single nucleus
in equilibrium with a gas of neutrons, protons, electrons and
$\alpha$-clusters.  The limit of these approaches is that they only consider $\alpha$ particles,
while many other nuclear species are expected to contribute to the equilibrium. 

The nuclear statistical equilibrium (NSE)
models \cite{raduta,hempel} go beyond SNA, because they consider all possible nuclear species in statistical equilibrium. However, in the original formulation of the approach \cite{iliadis2007}, the system is considered as an ideal gas of clusters, and nuclear interactions of the clusters among themselves, as well as with the surrounding gas of free nucleons, are neglected. As a result, the expected cluster melting at high density is not observed \cite{typel10}, showing that in-medium effects must be introduced. 
In the Hempel and Schaffner-Bielich \cite{hempel} model, such effects are  included within a geometric excluded-volume mechanism.
A more complex isospin dependence is proposed in the Raduta and Gulminelli model \cite{raduta2},
where an excluded volume-like correction is derived as a mass shift from the extended Thomas-Fermi
energy density functional \cite{Aymard14}. Still, all these approaches are semi-classical and do not account for temperature effects, which might explain why they only qualitatively agree with more microscopic treatments \cite{hempel-typel}. 
In particular, the excluded volume mechanism appears to provide a realistic treatment only at high densities close to the cluster dissolution density \cite{pais-typel}.
A better way to describe light clusters is the quantum statistical (QS) approach \cite{Roepke15} that can describe quantum correlations with the
medium, and takes into account the excited states and temperature effect. However, the mass shifts calculated within the QS approach are available only for a few nuclear species and a limited density domain, therefore they can be implemented in a complete NSE description of stellar matter only within some approximations \cite{typel16}.

A different approach was developed within the relativistic mean-field
framework and  uses mean-field concepts, such as the ones
 used in recent works \cite{typel10,ferreira12,pais15,avancini17},
 where light clusters are considered as new degrees of freedom. 
As such, they are characterized by a density, and possibly
  temperature, dependent effective mass, and they interact with the medium via meson couplings.
In-medium effects can thus be incorporated via the meson couplings, 
the effective mass shift, or both.
In Ref. \cite{typel10} the description of light clusters was achieved with a
 modification of the effective mass, which introduces a density and
 temperature dependence of the binding energy of the clusters. These
 quantities have been fitted to quantum statistical outputs for light
 clusters in warm matter. The
 meson-cluster couplings were taken proportional to the atomic number
 of the cluster, taking as basis the meson-nucleon couplings. 
However, it is not mandatory that the nucleons within the cluster feel the same mean field
as free nucleons.
 In Ref. \cite{avancini17}, the main idea was to obtain adequate phenomenological
 parametrizations for the clusters-mesons couplings. In particular, the
 authors looked for the parametrizations that better
 describe both experimentally obtained chemical equilibrium
 constants for  the formation of light clusters in 
 heavy-ion collisions \cite{qin12}, and microscopic results obtained from quantum
 statistical calculations. It was one goal of the work  \cite{avancini17}
  to discuss the combination of light cluster approaches with pasta structure concepts,
  which are important if going to high densities. As pointed out there, the coupling 
  to the  isoscalar-vector field was renormalized by a global parameter $\eta$
  to keep the parameter space restricted, but one should consider different couplings
  for the different clusters in future work to optimize the description of measured data,
  such as chemical equilibrium constants. 
In particular, whereas the $\alpha$ particles are well described by a suitable fit of $\eta$, 
the chemical equilibrium constants of the other light elements are not well reproduced.

To progress on a satisfactory description of light cluster degrees of freedom within the RMF framework,
in this article we explore the possibility of both in-medium mass shifts and in-medium modification of the cluster couplings. 
We aim at obtaining an universal, though phenomenological, set for the clusters-mesons couplings, 
with the purpose of having a formalism where different cluster species of arbitrary mass and charge can be described.  The inclusion of heavier clusters, i.e. pasta phases, will be left for a future work. 

At very low densities, a model independent constraint can be
considered: this is the one set by the virial EoS (VEoS) \cite{virial, typel12-v}, 
which only depends on the experimentally determined binding energies and 
scattering phase shifts, and provides the correct zero density limit for the equation of state 
at finite temperature. We therefore
fix the cluster-meson couplings so that, at very low densities, the  VEoS
particle fractions obtained in Ref.  \cite{typel12-v} are well
reproduced. 
The deuteron, which is weakly bound, needs a special treatment 
and will be considered later on.
We know that the VEoS breaks down when the interactions
between particles become stronger as the density increases. In this regime,
we use the fact that the cluster dissolution mechanism is reasonably well described 
by the geometrical excluded volume mechanism \cite{hempel-typel,pais-typel}, 
and employ the Thomas-Fermi formulation of Ref. \cite{raduta2} in order 
to evaluate the associated cluster mass shift. 
 
The final result is a simple analytical formula for the effective mass shift. 
To reproduce empirical data, an in-medium modified
coupling of cluster $j$ with the scalar meson $\sigma$  of the form $g_{sj}=x_s A_j g_s$ 
is proposed,
where $g_s$ is the coupling constant with the nucleons ($n, p)$,  
$A_j$ the cluster mass number, and $x_s$ is a universal cluster coupling fraction, 
with an associated uncertainty.

\section{Formalism}
In this section, we present the model used in the rest of the paper and
discuss how the light clusters, which are considered as point like
particles,  are included within our approach.

\subsection{Lagrangian}

Our system includes light clusters, both bosons, deuterons ($d,\,^2$H) and $\alpha$-particles ($^4$He), 
and fermions, tritons ($t,\,^3$H) and helions ($h,\,^3$He). 
They are immersed in a gas of  neutrons ($n$) and protons ($p$), neutralized by electrons.
The Lagrangian density of our system reads \cite{typel10,ferreira12,pais15,avancini17}:
\begin{eqnarray}
{\cal L} &=& \sum_{j=n,p,d,t,h,\alpha} {\cal L}_{j}                            
+ {\cal L}_{\sigma} + {\cal L}_{\omega} + {\cal L}_{\rho}+ {\cal L}_{\omega\rho}.
\end{eqnarray}
In the following, the couplings of the clusters to the mesons are
defined in terms of the couplings $g_s,\, g_v,\, g_\rho$ of the
nucleons to, respectively,  the $\sigma$, $\omega$ and
$\rho$-mesons. Besides, we will take for  the vacuum proton and
neutron mass an average value, $m=939$ MeV.
For the fermionic clusters,  $j=t,h$, we have:
\begin{eqnarray}
{\cal L}_j &=& \bar{\psi}\left[\gamma_\mu i D_j^\mu - M_j^*\right]\psi,
\end{eqnarray}
with  
\begin{equation}
iD^{\mu }_j = i \partial ^{\mu }-g_{vj} \omega^{\mu }-\frac{g_\rho}{2}{\boldsymbol\tau}_j \cdot \mathbf{b}^\mu ,
\end{equation}
where ${\boldsymbol \tau}_j$ is the isospin operator and $g_{vj}$ is the
  coupling of cluster $j$ to the vector meson $\omega$ and, in the
  present work, it is defined
  as $g_{vj}=A_j g_v$ for all clusters. The effective mass $M_j^*$ will be defined in the next section.

The Lagrangian density for the bosonic clusters, $j=d,\alpha$, is given by
\begin{eqnarray}
\mathcal{L}_{\alpha }&=&\frac{1}{2} (i D^{\mu}_{\alpha} \phi_{\alpha})^*
(i D_{\mu \alpha} \phi_{\alpha})-\frac{1}{2}\phi_{\alpha}^* \pc{M_{\alpha}^*}^2
\phi_{\alpha},\\
\mathcal{L}_{d}&=&\frac{1}{4} (i D^{\mu}_{d} \phi^{\nu}_{d}-
i D^{\nu}_{d} \phi^{\mu}_{d})^*
(i D_{d\mu} \phi_{d\nu}-i D_{d\nu} \phi_{d\mu})\nonumber\\
&&-\frac{1}{2}\phi^{\mu *}_{d} \pc{M_{d}^*}^2 \phi_{d\mu},
\end{eqnarray}
with
\begin{equation}
iD^{\mu }_j = i \partial ^{\mu }-g_{vj} \omega^{\mu }
\end{equation}

For the nucleonic gas,  $j=n,p$, we have:
\begin{eqnarray}
{\cal L}_j &=& \bar{\psi}\left[\gamma_\mu i D^\mu - m^*\right]\psi
\end{eqnarray}
with
\begin{eqnarray}
i D^\mu&=&i\partial^\mu-g_v\omega^\mu-\frac{g_\rho}{2}{\boldsymbol\tau}_j \cdot \mathbf{b}^\mu \\
m^*&=&m-g_s\phi_0 
\end{eqnarray}

For the fields, we have the standard RMF expressions:
\begin{eqnarray}
{\cal L}_\sigma&=&+\frac{1}{2}\left(\partial_{\mu}\phi\partial^{\mu}\phi
-m_s^2 \phi^2 - \frac{1}{3}\kappa \phi^3 -\frac{1}{12}\lambda\phi^4\right),\nonumber\\
{\cal L}_\omega&=&-\frac{1}{4}\Omega_{\mu\nu}\Omega^{\mu\nu}+\frac{1}{2}
m_v^2 V_{\mu}V^{\mu}, \nonumber \\ 
{\cal L}_\rho&=&-\frac{1}{4}\mathbf B_{\mu\nu}\cdot\mathbf B^{\mu\nu}+\frac{1}{2}
m_\rho^2 \mathbf b_{\mu}\cdot \mathbf b^{\mu}, \nonumber \\ 
{\cal L}_{\omega\rho}&=& g_{\omega\rho} g_\rho^2 g_v^2 V_{\mu}V^{\mu}\mathbf b_{\nu}\cdot \mathbf b^{\nu}
\end{eqnarray}
where
$\Omega_{\mu\nu}=\partial_{\mu}V_{\nu}-\partial_{\nu}V_{\mu}, $ and $ \mathbf B_{\mu\nu}=\partial_{\mu}\mathbf b_{\nu}-\partial_{\nu} \mathbf b_{\mu}
- g_\rho (\mathbf b_\mu \times \mathbf b_\nu)$.

\subsection{Mass shift in the clusters}

The total binding energy of a light cluster $j$ is given by 
\begin{eqnarray}
B_j=A_j m^*-M_j^* \,, \quad j=d,t,h,\alpha \,, \label{binding}
\end{eqnarray}
with $M_j^*$ the effective mass of cluster $j$, which is determined by
the meson coupling  as well as by a binding energy shift: 
\begin{eqnarray}
M_j^*&=&A_j m - g_{sj}\phi_0 - \left(B_j^0 + \delta B_j\right),
\label{meffi2}
\end{eqnarray}
Within the RMF approach, the nucleons are considered as independent moving particles, 
neglecting any correlations. The account of correlations via the introduction of bound states (clusters) 
will modify the coupling to the mesonic fields within the effective Lagrangian as denoted by the coupling constants $g_{sj}$ and $g_{vj}$.
There is no reason to consider them as the sum of the couplings of the individual constituents of the cluster, but 
they have to be introduced as new empirical parameters which are fitted to results from microscopic theories or to 
measured data. We discuss the choice of the coupling constants $g_{sj}$ in the following 
Section \ref{sec:results}, see also Eqs. (\ref{gs}) and (\ref{gv}) below.
In expression (\ref{meffi2}), $B^0_j$ is the binding energy of the cluster in the
vacuum and these constants are fixed to experimental values.
Following the formalism of Ref. \cite{Aymard14,raduta2}, we write for the binding energy shift  $\delta B_j$ 
\begin{eqnarray}
\delta
  B_j=\frac{Z_j}{\rho_0}\left(\epsilon_p^*-m\rho_p^*\right)+\frac{N_j}{\rho_0}\left(\epsilon_n^*-m\rho_n^*\right)
  \, ,
\label{deltaB}
\end{eqnarray}
which  is the energetic counterpart of the excluded volume mechanism in the Thomas-Fermi approximation. Here, $\rho_0$ is the nuclear saturation density. The energy states already occupied by the gas are excluded in the calculation of the cluster binding energy, thus avoiding double counting of the
particles of the gas and the ones of the clusters. 
The energy density, $\epsilon_j^*$,  and the density, $\rho_j^*$,  are
given by
\begin{eqnarray}
\epsilon_j^*&=&\frac{1}{\pi^2}\int_0^{p_{F_j}(\rm gas)} p^2 e_j(p) (f_{j+}(p)+f_{j-}(p)) dp \\
\rho_j^* &=&\frac{1}{\pi^2}\int_{0}^{p_{F_j}(\rm gas)}  p^2 (f_{j+}(p)+f_{j-}(p)) dp \,,
\end{eqnarray}
for $j=p,n$, and correspond to the gas energy density and the gas
  nucleonic density associated with the gas lowest energy levels. 
  In the last expressions, $f_{j\pm}$ are the usual Fermi
distribution functions for the nucleons and respective anti-particles, 
$p_{F_j}$ is the Fermi momentum of nucleon $j$, given by
  $p_{F_j}=(3\pi^2\rho_j)^{1/3}$,  and $e_j(p)=\sqrt{p_j^2+m^{*2}}$ is
  the corresponding
  single-particle energy of the nucleon $j$.

We treat the binding energy shifts, $\delta B_j$, as in Ref. \cite{typel10}: we replace the density dependence of these quantities by a vector meson dependence. This is equivalent, in our present study,  to consider in the shifts $\delta B_j$ the neutron and proton density replaced by

$$\rho_n=\frac{m_v^2}{2g_v} V_0-\frac{m_\rho^2}{2g_\rho}b_0, \qquad  \rho_p=\frac{m_v^2}{2g_v} V_0+\frac{m_\rho^2}{2g_\rho}b_0.$$ 

With the inclusion of this extra term, the equations for the fields read: 
\begin{eqnarray}
\label{field1}
&&m_{\rho, \rm eff}^2 b_0=\frac{g_{\rho}}{2}(\rho_p-\rho_n+\rho_{h}-\rho_t)  \\
&&-\frac{m_{\rho}^2}{g_{\rho}\rho_0}
\left(-\frac{\partial \epsilon^*}{\partial\rho_n}+\frac{\partial \epsilon^*}{\partial\rho_p}
+\frac{m\partial \rho^*}{\partial\rho_n}-\frac{m\partial \rho^*}{\partial\rho_p}\right)
\sum_j A_j \rho_s^j \nonumber \, ,   
\\
&&m_{v,  \rm eff}^2 V_0=g_{v}(\rho_p+\rho_n)+\sum_j g_{vj}\rho_{j}  \\ 
&&-\frac{m_v^2}{2g_v^2\rho_0}
\left(-\frac{\partial \epsilon^*}{\partial\rho_n}-\frac{\partial \epsilon^*}{\partial\rho_p}
+\frac{m\partial \rho^*}{\partial\rho_n}+\frac{m\partial \rho^*}{\partial\rho_p}\right)
\sum_j A_j \rho_s^j \nonumber \, ,   \label{field2}
 \\
&&m_s^2\phi_0 +\frac{k}{2}\phi_0^2+\frac{\lambda}{6}\phi_0^3=g_s\left(\rho_s^p+\rho_s^n\right) +\sum_j g_{sj}\rho_s^{j}\, ,  \label{field3}
\end{eqnarray}
with   $\epsilon^*=\epsilon_p^*+\epsilon_n^* \,,$  $\rho^*=\rho_p^*+\rho_n^* \,,$
  and 
\begin{eqnarray}
m_{\rho, \rm eff}^2&=&m_{\rho}^2+2g_{\omega\rho}g_{\rho}^2 g_v^2 V_0^2 \\
m_{v,  \rm eff}^2&=&m_{v}^2+2g_{\omega\rho}g_{\rho}^2 g_v^2 b_0^2 + \frac{1}{6}\xi g_v^4 V_0^2 \, .
\end{eqnarray}
For a given baryonic density, proton fraction and temperature, 
Eqs.~(\ref{field1}) - (\ref{field3}) have to be solved self-consistently.

\section{Results}
\label{sec:results}

\begin{figure*}[!htbp]
  \begin{tabular}{ccc}
\hspace*{-1cm}
\includegraphics[width=0.6\linewidth]{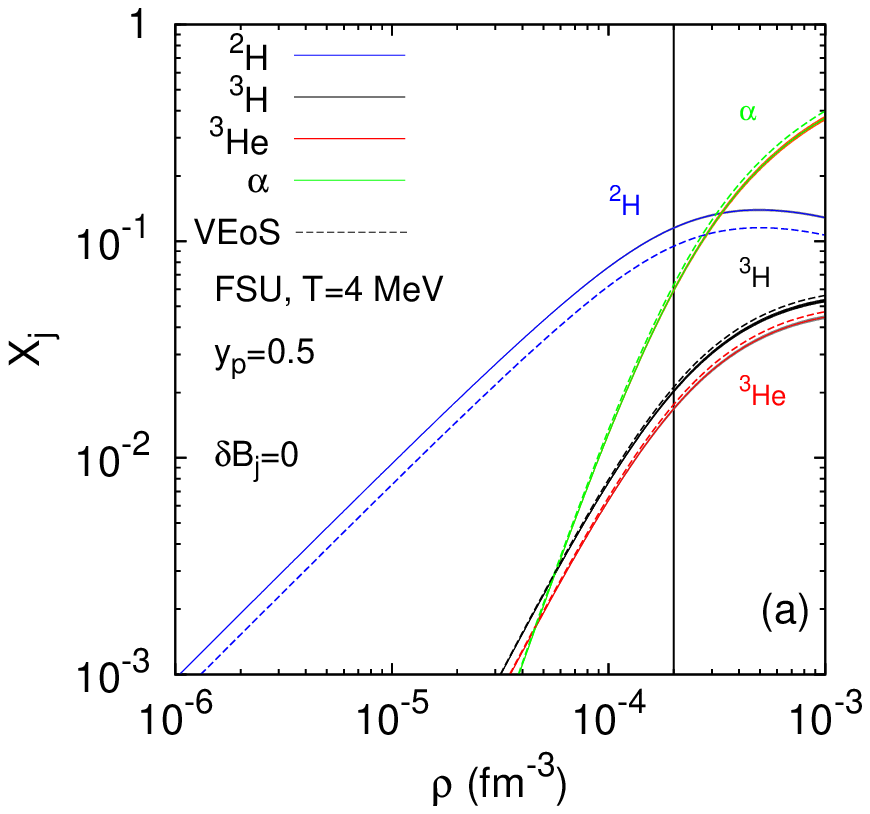} & \hspace*{-3cm}
\includegraphics[width=0.6\linewidth]{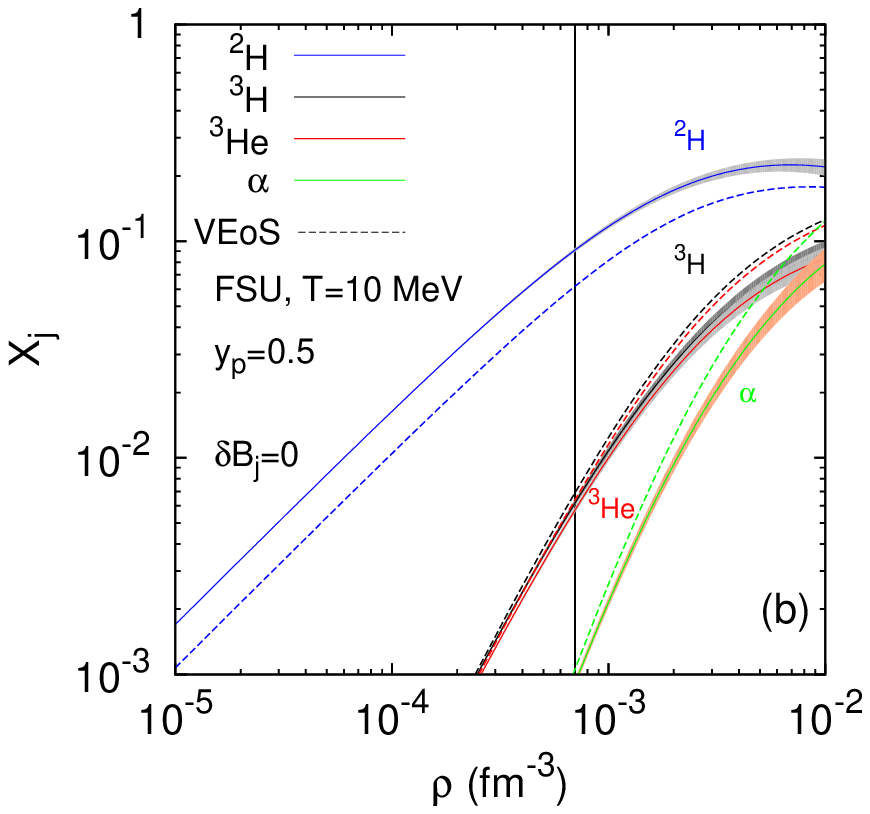}\\
 \end{tabular}
 \caption{(Color online) Fraction of deuteron, $X_d$,  triton, $X_{t}$, helion,
   $X_{h}$, and $\alpha$, $X_\alpha$, as a function of the density for FSU,
   $T=4$ MeV (a) and 10 MeV (b), with proton fraction $y_p=0.5$,
   taking $\delta B=0$,  $x_{sj}=0.85\pm 0.05$, (variation indicated by the spreading of the bands), and comparing with
   results of the Virial EoS from \cite{typel12-v}. Solid vertical black lines are given by {$\rho \lambda_n^3 =1/10$.} 
For more details, check the text.} 
\label{fig1}
\end{figure*}

In the following, we look for a possible
universal parametrization for all clusters which only account for the
differences through the atomic number and isospin projection. In the
last section, we test the proposed parametrizations by comparing the
predicted chemical equilibrium constants with the recent experimental
results published in Ref. \cite{qin12}.  All the calculations are  performed for the FSU
\cite{FSU} model, at finite  fixed temperatures and for fixed proton
fractions $y_p$ which describes the ratio of the total proton density 
to the baryon density. For this model, the values of the
  nucleon coupling constants are $g_s^2=112.1996, g_v^2=204.5469$, and
  $g_\rho^2=138.4701$ and the nuclear saturation density is $\rho_0=0.148 \,{\rm fm}^{-3}$.
 Further constants (the meson masses and the couplings of the
  non-linear meson terms) are found in Ref. \cite{FSU}.  
 This model has been chosen because it describes adequately the
  properties of nuclear matter at saturation and subsaturation densities. It has
  the drawback of not predicting a two solar mass NS. However,  it is
  possible to include excluded volume like
  effects above saturation density making the EoS hard enough at high
  density \cite{maslov15,dutra16}. We have tested the formalism with two other models that
  have good properties at saturation density and below, and, besides,
  describe two solar mass NS, the NL3$\omega\rho$  \cite{horowitz01} and the
  TM1$\omega\rho$ models \cite{providencia13,bao14} with the symmetry energy slope $L\sim 55$
  MeV. The results obtained were within the uncertainty bands of our approach and, therefore, we do
  not include them in the present study. A more complete thermodynamical study  will be left for a future work.

\subsection{Low-density limit and cluster-meson couplings} 
\label{sec:Cluster-meson Couplings}

\begin{table}[!htbp]
\caption{Virial cluster fraction, $X_j$, for the light 
  clusters triton, helion and $\alpha$, at different densities, $\rho$, for
  $T=4$  and 10 MeV used in the present work and taken from Ref. \cite{typel12-v}. The densities are in units of 10$^{-6}$fm$^{-3}$.}
   \begin{tabular}{lccccc}
\hline
$\rho$ 	&1.1&	5.3&	12.0&	52.5  &91.2\\ 
& \multicolumn{5}{c}{$X_j$}\\
\hline
cluster &\multicolumn{5}{c}{$T=4$ MeV}\\
$t (^3$H)&1.3$\times 10^{-6}$&3.0$\times 10^{-5}$&1.5$\times 10^{-4}$&2.6$\times 10^{-3}$&6.8$\times 10^{-3}$\\
$h (^{3}$He)&1.1$\times 10^{-6}$&2.5$\times 10^{-5}$&1.3$\times 10^{-4}$&2.1$\times 10^{-3}$&5.7$\times 10^{-3}$\\
$\alpha (^{4}$He) &2.7$\times 10^{-8}$&2.9$\times 10^{-6}$&3.2$\times 10^{-5}$&2.4$\times 10^{-3}$&1.1$\times 10^{-1}$\\
\hline
cluster &\multicolumn{5}{c}{$T=10$ MeV}\\
$t (^3$H)&2.3$\times 10^{-8}$&5.2$\times 10^{-7}$&2.7$\times 10^{-6}$&5.1$\times 10^{-5}$&1.5$\times 10^{-4}$\\
$h (^{3}$He)&2.1$\times 10^{-8}$&4.8$\times 10^{-7}$&2.5$\times 10^{-6}$&24.7$\times 10^{-5}$&1.4$\times 10^{-4}$\\
$\alpha (^{4}$He) &6.0$\times 10^{-12}$&6.7$\times 10^{-10}$&7.9$\times
                                                     10^{-9}$&6.4$\times
                                                               10^{-8}$&
   2.3$\times 10^{-6}$\\
\hline
\end{tabular}
\label{tab1}
\end{table}

We will first take as reference the  virial EoS (VEoS)
\cite{typel12-v}. There, the account of continuum correlations (scattering phase shifts), which is necessary 
to obtain the correct second virial coefficient, was performed by introducing a temperature dependent 
effective resonance energy $E_{ij}(T)$ in each $ij$ channel.
The cluster-meson couplings  are obtained
from the best fit of the RMF cluster fractions, defined as $X_j = A_j n_j/n$, to these data, taking  the FSU parametrization. 
The fit is done choosing a sufficiently low density close to the cluster
onset where the virial EoS is still valid and at the same time the interaction already has
non-negligible effects,  see Table \ref{tab1}. 
We have considered densities  between $10^{-6}$ fm$^{-3}$ and  $10^{-4}$ fm$^{-3}$, though, for small temperatures, $10^{-4}$ fm$^{-3}$ is close to  the limit of validity of the VEoS. 
{Still,} we expect that at these densities the VEoS is a
good approximation.  
In this low density domain, the binding energy shift  $\delta
B_j$ of Eq. (\ref{deltaB}) is completely negligible and does not affect the particle fractions
(see also Figure \ref{fig2} below), therefore it was put to zero for this calculation. However, 
already at $5\times 10^{-6}$ fm$^{-3}$, the cluster fractions are sensitive 
to the meson couplings. 

In principle, both scalar and vector couplings could be considered for
this fit. However,  for the presently
existing constraints, it was shown in Ref. \cite{ferreira12} that the
$\{g_{sj},g_{vj}\}$ parameter space is somewhat redundant and very
similar results can be obtained either by modifying the scalar
coupling (i.e. decreasing the nuclear attraction) or the vector one
(i.e. increasing the nuclear repulsion).  
In contrast to Ref. \cite{ferreira12} where $g_v$ was scaled, in this work we
only optimize the $g_{sj}$ parameters,
\begin{equation}
\label{gs}
g_{sj}=x_{sj} A_j g_s,
\end{equation}
 while the vector couplings are
set to 
\begin{equation}
\label{gv}
g_{vj}=A_j g_v.
\end{equation}

 We have performed calculations for
$T=4$ and 10 MeV, keeping the proton fraction at 0.5. 

 It is clear that
we are not able to reproduce the deuteron fractions predicted by the
VEoS. This is somewhat expected, due to the specificity of the deuteron. 
Indeed such a loosely bound structure which is known to correspond to highly 
delocalized wave function can be hardly described in a mean-field approximation.
As detailed in Ref.~\cite{Roepke15}, if the binding energy per nucleon
is small compared with $T$, the contributions of the continuum as given 
by the scattering phase shifts are essential. For the other clusters our coupling parametrizations are
reasonable within the range of temperatures between 4 and 10 MeV. 

Reasonable values for
$g_{sj}$ are $(0.85\pm 0.05) A_j g_s$, see Fig. \ref{fig1},  where the colored
bands show the range of particle fractions  covered by this interval
at low densities, for $T=$4 and 10 MeV.  The solid vertical
  black lines represent the upper limit of the region of validity of the VEoS. This region can be estimated imposing $\rho_j
  \lambda_j^3 << 1$, where  $\lambda_j=\sqrt{2\pi/(m_j T)}$ is the
  thermal wavelength of particle $j$, $\rho_j$ its density and $m_j$
  its mass \cite{typel12-v}. The vertical line was defined by $\rho
  \lambda_n^3 = 1/10$, $\rho$ being the baryon density and
  $\lambda_n$ the nucleon thermal wavelength. 
In Table \ref{tab2}, we compare explicitly the RMF abundances of clusters  obtained
under these conditions with the ones coming from the VEoS. 
Different values for the clusters $\sigma$-meson couplings,
$g_{sj}$, Eq. (\ref{gs}), were considered  for five different values
for the density.  In particular, we have taken for the fraction
$x_{sj}$ three values which best describe the VEoS. 
For the lowest density in Table \ref{tab2},
the cluster fractions are almost independent of
$x_{sj}$ because for this low density  the clusters behave like free
particles. At $T=10$ ($T=4$) MeV the largest deviations obtained are
below 2\% (5\%)
of relative difference for the largest density considered, the largest
deviations occurring for the $\alpha$-clusters. Choosing the best
couplings, these relative differences can be reduced to $\sim 1\%$
($\sim 2\%$).

Larger values of $x_{sj}$ were considered  but we found
the problem already discussed in Ref.
\cite{ferreira12}: taking 
$g_{vj}=A_j g_v$, the light clusters will not dissolve if $x_{sj} \ge 1$.
We have confirmed that even including the contribution
$\delta B_j$, the clusters would not dissolve with this value of
$x_{sj}$.

\begin{table}[!htbp]
\caption{\label{tab2}
Relative difference in percentage of the cluster fractions between the VEoS and the
  RMF EoS [$\Delta_{\rm rel}=100\times (X_{j}^{\rm RMF}-X_j^{\rm VEoS})/ X_j^{\rm VEoS}]$
  for different couplings, $g_{sj}=x_{sj}A_j g_s$ and $g_{vj}=A_j g_v$, 
and baryon number densities $\rho$ for the light   clusters triton ($t$), 
helion ($h$), and $\alpha$,  with $T=4$  and 10 MeV.  The densities are in units of 10$^{-6}$fm$^{-3}$.}
 \begin{tabular}{lccccc}
\hline
$\rho$ 	&1.1&	5.3&	12.0&	52.5
        &91.2\\ 
& \multicolumn{5}{c}{$\Delta_{\rm rel}(\%)$ }\\
\hline
$x_{sj}$&\multicolumn{5}{c}{$T=4$ MeV}\\
\hline
triton ($t$) \\
0.80	&0.24	&0.01	&-0.37	&-2.33	&-3.76\\
0.85	&0.26	&0.07	&-0.22	&-1.74	&-2.85\\
0.90	&0.27	&0.14	&-0.06	&-1.13	&-1.92\\
helion ($h$)\\
0.80	&0.11	&-0.13	&-0.36	&-2.48	&-3.91\\
0.85	&0.12	&-0.06	&-0.36	&-1.88	&-2.99\\
0.90	&0.14	&0.01	&-0.20	&-1.28	&-2.06\\
$\alpha$\\
0.80	&0.07	&-0.24	&-0.74	&-3.35	&-5.22\\
0.85	&0.09	&-0.15	&-0.53	&-2.55	&-4.02\\
0.90	&0.11	&-0.06	&-0.33	&-1.76	&-2.78\\
\hline
\multicolumn{6}{c}{$T=10$ MeV}\\
\hline
triton ($t$)\\
0.80	&0.62	&0.57	&0.39	&-0.67	&-1.64\\
0.85	&0.62	&0.59	&0.45	&-0.40	&-1.18\\
0.90	&0.63	&0.62	&0.51	&-0.13	&-0.73\\
helion ($h$)\\
0.80	&0.48	&0.43	&0.25	&-0.81	&-1.78\\
0.85	&0.49	&0.46	&0.31	&-0.54	&-1.32\\
0.90	&0.49	&0.49	&0.38	&-0.27	&-0.87\\
$\alpha$\\
0.80	&0.34	&0.27	&0.03	&-1.36	&-2.64\\
0.85	&0.34	&0.31	&0.12	&-1.01	&-2.03\\
0.90	&0.35	&0.35	&0.20	&-0.65	&-1.43\\
\hline
\end{tabular}
\end{table}

Next we will discuss the effect of introducing a non-zero binding energy
shift $\delta B_j$,
Eq.~(\ref{deltaB}).
In Fig. \ref{fig2},
we compare the binding energy
of the $\alpha$-clusters obtained taking  $\delta
B_{\alpha}$  defined by Eq. (\ref{deltaB})  with the binding
energy
\begin{equation}
B_j=B_j^0+\delta B_j^{\rm QS}
\label{bindTypel}
\end{equation}
 obtained from QS calculations. In particular, a perturbation theory was
given in Ref.
\cite{Roepke09}, where the following result
for the Pauli blocking shift of $\alpha$ particles with center-of-mass
momentum
(wave number) $P = 0$
\begin{equation}
\label{perturbation}
 \delta B_\alpha^{\rm
Pauli}(P=0;\rho_n,\rho_p,T)=-\frac{164371\,\,\rho}{(T+10.67)^{3/2}}
\end{equation}
(in units MeV, fm) was obtained in lowest order of density $\rho$.
Typel et al. \cite{typel10} performed RMF calculations for a wide
density region.
To suppress cluster formation at higher densities, an empirical
quadratic form
was introduced,
\begin{eqnarray}
\label{Typel}
 &&\delta B_\alpha^{\rm Typel}(T)=\delta B_\alpha^{\rm
Pauli}(P=0;\rho_n,\rho_p,T)
\nonumber \\ && \times
\left[1-\frac{\delta B_\alpha^{\rm Pauli}(P=0;\rho_n,\rho_p,T)}{2
B_\alpha^0}\right].
\end{eqnarray}

Both results are shown in Fig.~\ref{fig2}, together with more recent
calculations for 
$\delta B_j^{\rm QS}(P;\rho_n,\rho_p,T)$ according to Ref.~\cite{Roepke15}.
In contrast to the Pauli blocking assuming the ideal Fermi distribution
in the nuclear medium,
correlations in the medium have been taken into account, and the
distribution function
of the nucleons in the medium is parametrized there by a Fermi
distribution with effective chemical
potentials and temperature. Also shown in Fig.~\ref{fig2} are QS
calculations for different
center-of-mass momenta $P=0, 1, 2$ fm$^{-1}$.

\begin{figure}[!htbp]
 \begin{tabular}{cc}
\includegraphics[width=0.5\textwidth]{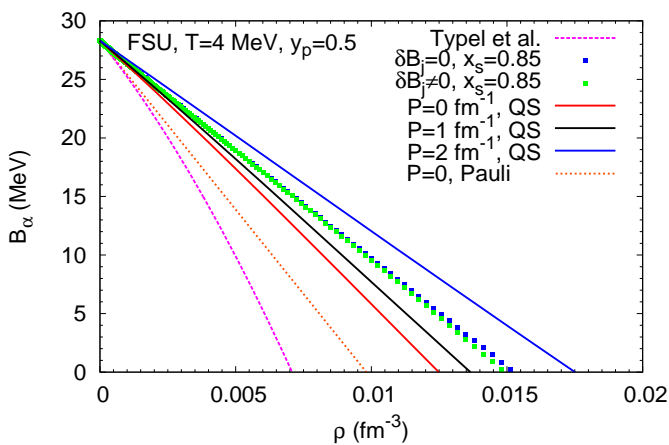}
 \end{tabular}
  \caption{(Color online) Binding energy of $\alpha$  for the RMF-FSU calculation (this
work), $T=4$ MeV, and
    $y_p=0.5$ obtained with Eq.~(\ref{meffi2}). For comparison, results
neglecting the
binding energy shift (\ref{deltaB}) ($\delta B_j=0$),
as well as QS calculations of a perturbative approach \cite{Roepke09} (Pauli),
Eq.~(\ref{perturbation}), the empirical form  Eq.~(\ref{Typel}) from 
Typel et al.
\cite{typel10}, and results obtained from a recent QS approach
\cite{Roepke15} for different
     center-of-mass momenta $P$ are also shown. }
\label{fig2}
\end{figure}

The QS calculations show two effects: \\
(i) due to the in-medium correlations the
perturbation theory result for the Mott density, according
to \cite{Roepke09},  is shifted to higher densities.\\
(ii) The Pauli blocking is strongly dependent on the center-of-mass
momentum $P$ of the cluster.
For the $\alpha$ cluster at temperatures and densities considered here,
typical values of $P$ for the
four-nucleon contribution to the EoS are of the order 1 fm$^{-1}$.

We can also see that the shifts given by the empirical
reduction of the coupling to the $\sigma$ meson field proposed in this
work to reproduce the VEoS, and the microscopically calculated shifts of the binding energies of the QS
calculations, are of the same order.
This is a prerequisite for the description of the composition and the chemical equilibrium constants as discussed in the following sections. 

\begin{figure}[!htbp]
 \begin{tabular}{cc}
\includegraphics[width=0.5\textwidth]{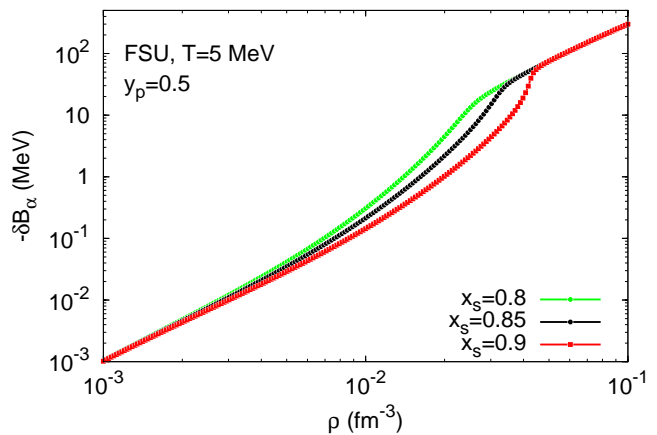} \\
\includegraphics[width=0.5\textwidth]{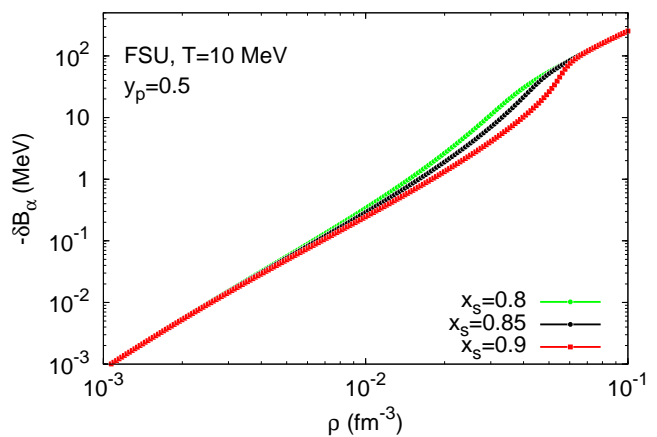} \\
\includegraphics[width=0.5\textwidth]{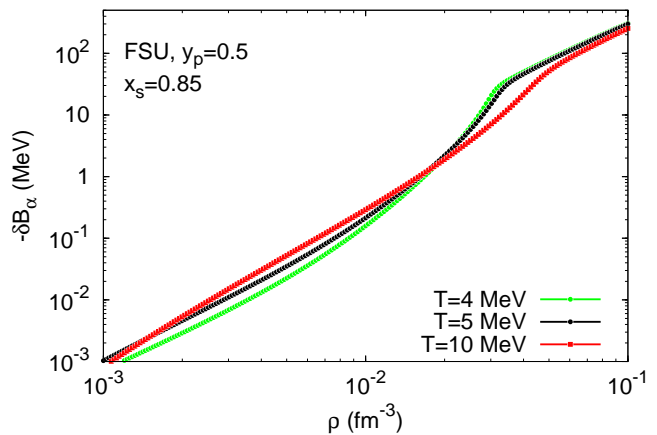} \\
 \end{tabular}
  \caption{(Color online) Binding energy shift of $\alpha$, $\delta B_\alpha$, as given by  Eq.(\ref{deltaB})  for the RMF-FSU calculation, $y_p=0.5$, $T=5$ MeV (top), $T=10$ MeV (middle), for $x_s=0.8,0.85,0.9$. The bottom panel shows the same shift for all the temperatures and keeping $x_s=0.85$. }
\label{fig3}
\end{figure}

\begin{figure*}[!htbp]
  \begin{tabular}{c}
\includegraphics[width=0.8\textwidth]{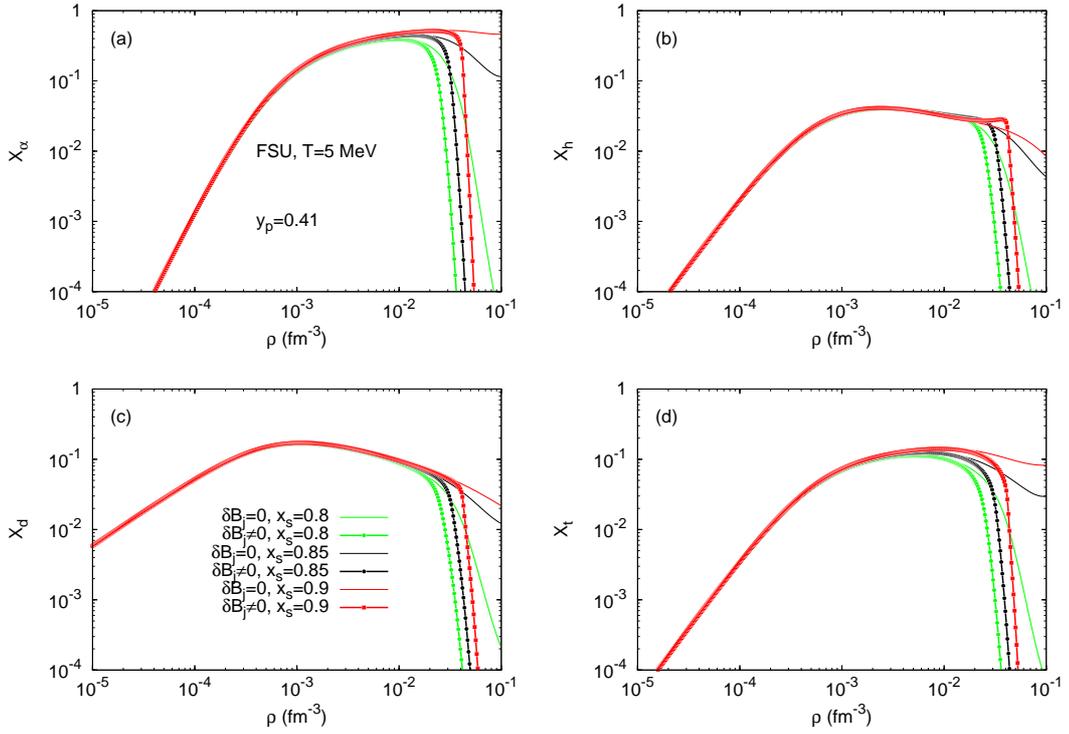} \\ 
\end{tabular}
 \caption{(Color online) Fraction of $\alpha$, $X_{\alpha}$ (a), helion, $X_{h}$ (b),  deuteron, $X_d$ (c), and triton, $X_t$ (d), as a function of the density for FSU, $T=5$ MeV,
   and $y_p=0.41$, with and without $\delta B_j$, for $x_{sj}=0.8, 0.85,
   0.9$, keeping $g_{vj}=A_j g_v$. 
     } 
\label{fig4}
\end{figure*}

The additional binding energy shift $\delta B_j$ given by Eq.~(\ref{deltaB})
is completely negligible
in the domain of validity of the VEoS, which means that the cluster
couplings extracted
in Table~\ref{tab2} do not depend on this term.
 Even at higher density, this extra correction is small
in the range of densities where the binding energies of the clusters are still positive, but rises fast for larger densities, see Fig. \ref{fig3}.
It gives an important contribution to the
  self-consistent calculation of matter in thermodynamic equilibrium
  at higher densities, as it will be shown in Figs. ~\ref{fig4},
  ~\ref{fig5} and ~\ref{fig6}. 

It is also interesting to discuss  the effect of the coupling
$x_{sj}$ and temperature $T$ on the binding energy shift. From
Figs. \ref{fig3} we conclude that  the larger $x_{sj}$ the slower 
{-$\delta B_j$} increases and also that a larger temperature determines a softer
behavior, with {-$\delta B_j$}  taking larger values at the lower densities and smaller ones close
to the dissolution density.

The change of slope for $\rho>0.02$ fm$^{-3}$ occurs at the maximum of the cluster fraction, corresponding to the onset the cluster fraction starts to decrease.

However, we should stress that
the representation of Figs. \ref{fig2} and \ref{fig3} do not give a complete picture
of the in-medium effects and cluster dissolution mechanism. As we can
see from Eqs. (\ref{field1}) - (\ref{field3}), the mass shift deeply
modifies the equations of motion for the meson fields. The particle
fractions are thus affected in a highly complex way because of the
self-consistency of the approach, which additionally induces temperature
effects.

\subsection{Global cluster distributions}

We are interested in extending the calculation of thermodynamic properties {from the low-density 
region where perturbation theory can be applied} to the entire 
subsaturation region $\rho \le \rho_0$. It is expected that the light clusters will be dissolved 
 below $\rho_0$, and the RMF approach is applicable there. Correlations which are always present in 
nuclear matter are included in this density-functional approach, and the fit to data at saturation 
density presumes that no further correlations are considered.

The description of the fractions of different components is difficult not only because of the 
problems with the many-body theory at high densities, 
but also the conceptual definitions of bound states near the Mott density is problematic. 
The QS theory has been worked out for the two-particle case and extrapolated for the other light
clusters, see Ref. \cite{Roepke15}. A further problem is that at higher densities also other 
structures such as pasta structures are of relevance so that one cannot discuss 
the thermodynamics at higher densities without the account of droplets
and other structures.

We show in the present subsection that the clusters are dissolved below $\rho_0$. 
This has been achieved, for instance, by Typel et al. \cite{typel10} introducing an empirical
quadratic form (\ref{Typel}). A more microscopic approach to the suppression of the cluster fraction
was given in Ref. \cite{Roepke15} where the dissolution of the bound states 
and the virial contributions for the partial partition functions are considered. In this work we
show that the account of the binding energy shift $\delta B_j$, Eq. (\ref{deltaB}), gives similar
results.

\begin{figure*}[!htbp]
  \begin{tabular}{c}
\includegraphics[width=0.8\textwidth]{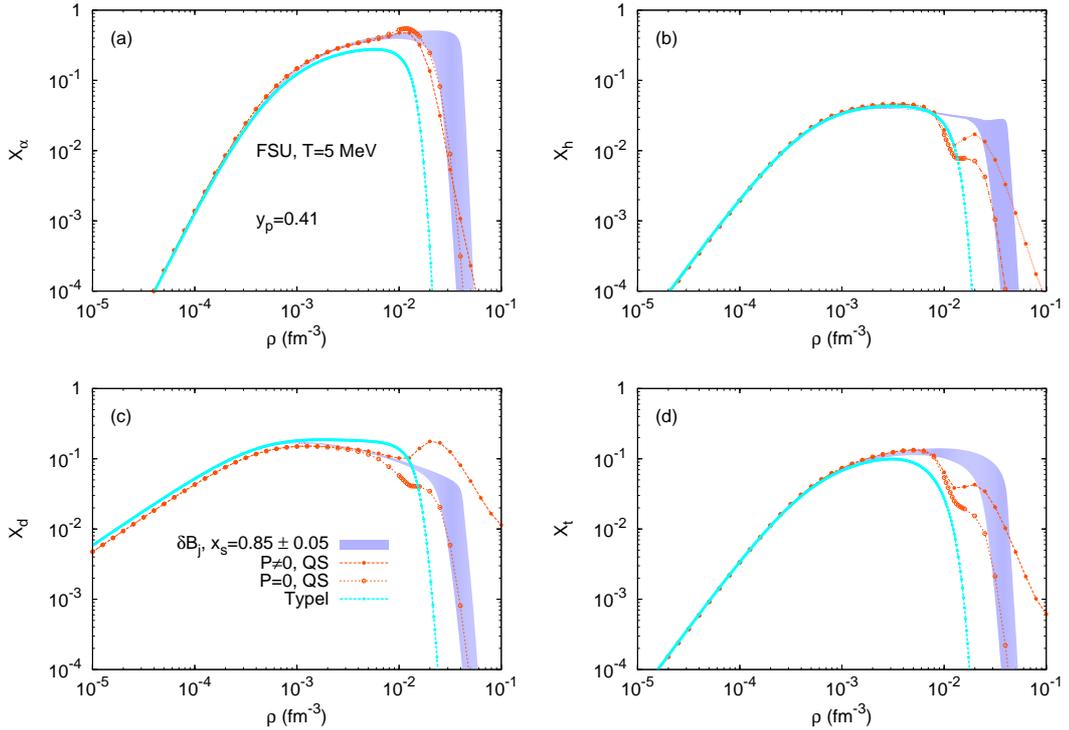} \\ 
\end{tabular}
 \caption{(Color online) Fraction of $\alpha$, $X_{\alpha}$ (a), helion, $X_{h}$ (b),  deuteron, $X_d$ (c), and triton, $X_t$ (d), as a function of the density for FSU, $T=5$ MeV,
   and $y_p=0.41$, with $\delta B_j$, for $0.8 <x_{sj}<
   0.9$, keeping $g_{vj}=A_j g_v$,  comparing with the QS EoS from \cite{Roepke15}
   with full $\bf P$ dependence of the residual virial coefficient (red line with full dots), and neglecting the 
dependence (${\bf P}=0$) (red line with empty dots), and with the EoS given by Typel et al. in \cite{typel10} (cyan full lines). 
     } 
\label{fig5}
\end{figure*}

\begin{figure*}[!htbp]
  \begin{tabular}{c}
\includegraphics[width=0.8\textwidth]{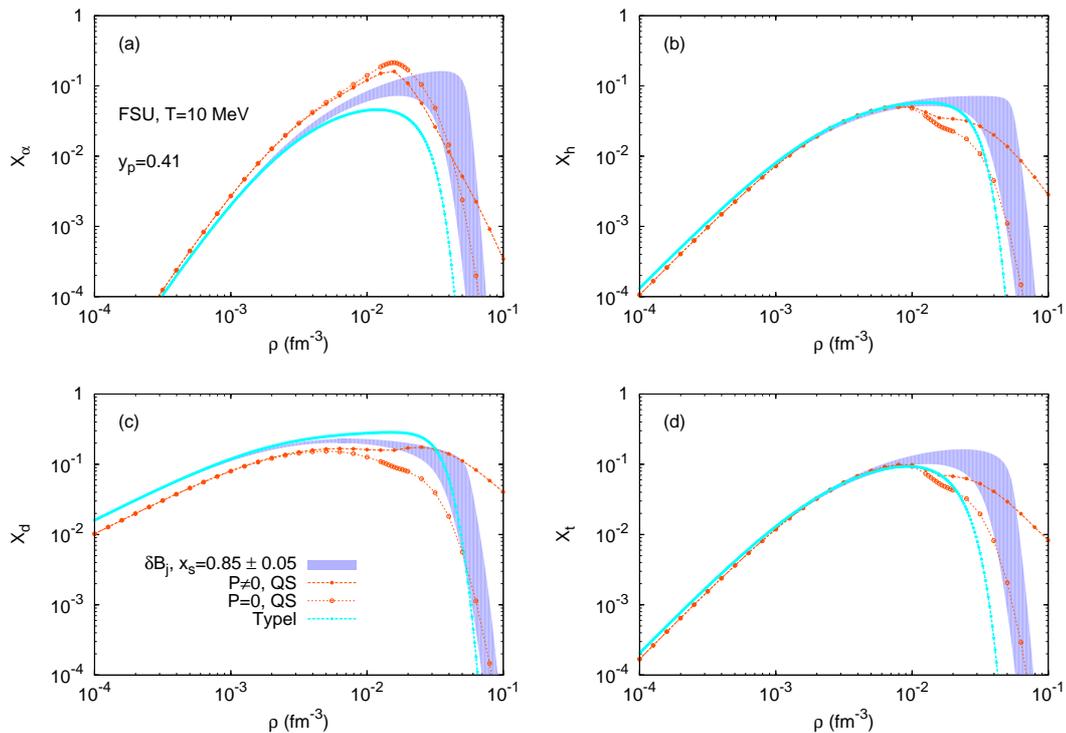} \\ 
 \end{tabular}
 \caption{(Color online) Fraction of $\alpha$, $X_{\alpha}$ (a), helion, $X_{h}$ (b),  deuteron, $X_d$ (c), and triton, $X_t$ (d), as a function of the density for FSU, $T=10$ MeV,
   and $y_p=0.41$, with $\delta B_j$, for $0.8 <x_{sj}<
   0.9$, keeping $g_{vj}=A_j g_v$,  comparing with the QS EoS from \cite{Roepke15}
   with full $\bf P$ dependence of the residual virial coefficient (red line with full dots), and neglecting the 
dependence (${\bf P}=0$) (red line with empty dots), and with the EoS given by Typel et al. in \cite{typel10} (cyan full lines). 
        }
\label{fig6}
\end{figure*}

The global effect of the modified meson couplings and binding energy shifts
is presented in Figs.~\ref{fig4}, ~\ref{fig5} and \ref{fig6}.

 In Fig.~\ref{fig4}, we show the clusters fractions for matter with a fixed proton fraction of $y_p=0.41$,
 and $T=5$ MeV,  keeping $g_{vj}=A_j g_v$ and using different
values for $g_{sj}=x_{sj} A_j g_s$: $x_{sj}=0.8,\, 0.85,\, 0.9$. We extend the
cluster fractions to larger densities in order to analyse the $\delta
B_j $ contribution. Neglecting this term, the clusters do not
dissolve.  Taking $x_{sj}=0.9$, the clusters seem to dissolve but there is just a local reduction of clusters
 followed by a reappearance to similar fractions. 
The role of the extra term in the binding energy is precisely to
dissolve the clusters at large densities,  and the larger the value of $x_{sj}$
 the larger the dissolution density. Typical values for the dissolution ($X_j < 10^{-4}$)
of light clusters  at conditions considered here are densities $0.04 \,{\rm fm}^{-3} < \rho 
< 0.06\, {\rm fm}^{-3}$.

To compare with QS calculations \cite{Roepke15}, we show in Fig.~\ref{fig5} the mass fractions
of light clusters for the parameter region $0.8 \le x_{sj} \le 0.9$ including 
the binding energy shift $\delta B_j$, Eq.~(\ref{deltaB}), represented by a band, together with the 
QS calculation  taking into account the 
Pauli blocking term in the residual virial coefficient $v_i({\bf P};T,\rho,y_p)$
as function of the center-of mass momentum $\bf P$, see Sec.~V.D. of Ref.~\cite{Roepke15}. 
Good agreement of both approaches is seen up to density $\rho \approx 0.01$ fm$^{-3}$.

The disappearance of the clusters at higher densities in the QS approach 
is not so sharp as expected from the assumption that, near the saturation density, 
 nuclear matter is fully described by the RMF approach in an empirical way, similar to a density functional approach. 
In principle, one has to analyse which microscopic correlations are always contained in this 
effective mean-field approach. These correlations have to be removed from the contribution of 
light clusters to the thermodynamic properties. Whereas this problem has been solved in the 
low-density limit, see \cite{Roepke15}, a rigorous solution analysing continuum correlations 
near the saturation density is not at reach yet. 

Two approaches to suppress the contribution of
clusters at high densities are shown in Fig.~\ref{fig5}: (i) as already discussed above, 
according to Typel et al. \cite{typel10} a quadratic term is introduced in an empirical way
to calculate the shift of the binding energy of clusters, Eq. (\ref{bindTypel}). The result
shown in Fig.~\ref{fig5} gives a suppression which is too strong compared with the other 
approaches, see also Fig.~\ref{fig2}; (ii) a stronger suppression of clusters at increasing density
is also obtained if the residual virial coefficient $v_i(0;T,\rho,y_p)$ is used, neglecting
the $\bf P$ dependence, as shown in \cite{Roepke15}. This result for ${\bf P} = 0$ is also shown in
 Fig.~\ref{fig5}. Good agreement with the RMF approach is obtained for the dissolution density,
but stronger deviations ("bumps") occur for densities $\rho > 0.01$ fm$^{-3}$. 
This is a consequence of the fact that the composition is strongly interdependent, an overproduction 
of $\alpha$ particles is connected with an underproduction of other clusters. 

In conclusion, the RMF approach considered here seems to be an appropriate description of the composition of nuclear matter also at high densities.

In Fig.~\ref{fig6},  the same analysis is done at 
$T=10$ MeV including the $\delta
B_j$ contribution. 
As in Fig.~\ref{fig5}, our results are compared to the fractions obtained with the binding energy shift
of Ref.~\cite{typel10} fitted on QS results, and the fully microscopic QS EoS 
of Ref.~\cite{Roepke15}.
We can see that the different models very well agree at low density. The only sizeable difference is a reduced deuteron fraction for the QS calculation, which is the only one to reproduce the VEoS for deuterons, as expected. At high density, the phenomenological models correctly obtain the cluster dissolution but the dissolution density is model dependent. This model dependence cannot be reduced using the QS microscopic results as a constraint, because these latter lack high order correlations at high density, where the different phenomenological prescriptions more strongly differ.

Concerning the temperature dependence of the cluster dissolution
mechanism, we consider the fraction of each  light cluster as $X_{\rm j}=10^{-4}$ to get the dissolution density, $\rho_{{\rm diss}}(T)$. 
Results for the different clusters are shown
 in Tab.~\ref{tab3}. We can quantify the temperature dependence of the cluster dissolution density
introducing a variable also shown in Tab.~\ref{tab3},
\begin{equation}
 \Gamma_{\rho_{{\rm diss}}}=\rho_{{\rm diss}}(T=10)/\rho_{{\rm diss}}(T=5)
\end{equation}
defined by the ratio, for each particle species $j$, of their
dissolution density at the higher temperature, and the dissolution
density at the lower temperature.
We can see that  the effect of temperature strongly depends
on the chosen coupling, the biggest effect being obtained with the
smallest value for the coupling $x_s$.
The binding energy shift of Eq. (\ref{deltaB}) leads to  $\Gamma_{\rho_{{\rm diss}}}(j)$
of the order of {1.4 (1.6)} for the $\alpha$ (deuteron) for $x_s=0.85$,  see Tab.
\ref{tab3}. Identical values are obtained choosing
$x_s=0.8$, while a slightly smaller temperature effect is seen with
$x_s=0.9$,  $\Gamma_{\rho_{{\rm diss}}}(j)\approx 1.4-1.5$. Similar results are determined from the QS calculations with $\bf P$ = 0 of Ref. \cite{Roepke15}. 
We also compare with the dissolution density  $\rho_{{\rm diss}}$
of Typel et al. \cite{typel10}, which shows a significantly larger temperature dependence of the clusters dissolution density expressed by $\Gamma_{\rho_{{\rm diss}}}$.

\begin{table}[!htbp]
\caption{\label{tab3}
Dissolution density, $\rho_{{\rm diss}}$, for each cluster, considering $x_s=0.8,0.85$, and 0.9, for $T=5$ and 10 with $y_p=0.41$.
We considered $X_{{\rm j}}=10^{-4}$ to get $\rho_{{\rm diss}}$. The last part of the table shows $\Gamma_{\rho_{{\rm diss}}}=\rho_{{\rm diss}}(T=10)/\rho_{{\rm diss}}(T=5)$. We also compare with the dissolution density of Typel et al. \cite{typel10}, and with the QS EoS from \cite{Roepke15}, neglecting the ${\bf P}$ dependence (${\bf P}=0$).
}
 \begin{tabular}{lccccc}
 \hline
\hline
&\multicolumn{4}{c}{$T=5$ MeV}\\
\hline
$\rho_{{\rm diss}}$	&$x_s=0.8$&$x_s=0.85$&$x_s=0.9$&\cite{typel10} &QS\\ 
\hline
deuteron		&0.04164	&0.04954	&0.05903	&0.02389	&0.04793\\
triton		&0.03615	&0.04374	&0.05295&0.01772	&0.04244\\
helion      &0.03577&0.04394&0.05404&0.01880 	&0.04004\\
$\alpha$	 	&0.03629	&0.04429	&0.05406&0.02113	& 0.04244\\
\hline
&\multicolumn{4}{c}{$T=10$ MeV}\\
\hline
$\rho_{{\rm diss}}$	&$x_s=0.8$&$x_s=0.85$&$x_s=0.9$&\cite{typel10} &QS\\ 
\hline
deuteron		&0.06984	&0.07983	&0.09106	&0.06490	&0.08221\\
triton		&0.05743&0.06706	&0.07801&0.04259	&0.06909\\
helion      &0.05794&0.06860&0.08084&0.04856 	&0.06515\\
$\alpha$	 	&0.05298&0.06328	&0.07511&0.04404	&0.06623\\
\hline
$\Gamma_{\rho_{{\rm diss}}}$	&$x_s=0.8$&$x_s=0.85$&$x_s=0.9$&\cite{typel10} &QS\\ 
\hline
deuteron		&1.6772	&1.6114	&1.5426&2.7166	&1.7153\\
triton		&1.5887&1.5331	&1.4733&2.4035	&1.6279\\
helion      &1.6198&1.5612&1.4904&2.5830 &1.6268\\
$\alpha$	 	&1.4599&1.4288	&1.3894&	2.0842	&1.5604\\
\hline
\hline
\end{tabular}
\end{table}

Again, if the qualitative effect of a dissolution density increasing with increasing temperature is physically reasonable and well understood \cite{typel10,Roepke15}, a quantitative determination is less obvious, and within the present constraints it is not easy
to discriminate between the different predictions.

From the model dependence point of view, we have seen that choosing
different coupling fractions for the different nuclear species, within
the constraint of the VEoS at low density, still produces abundances
which are within the uncertainty determined in Figs. \ref{fig5}, \ref{fig6} 
considering a universal coupling.

 This is true concerning both  the temperature and the density dependence. 
In the absence of more constraining observations/calculations, we can propose  $x_{s}=0.85  \pm 0.05$ as a reasonable universal value for the cluster couplings.

 Let us comment on the fact that we are only considering light
  clusters in the present study. Indeed, in-medium effects 
on the heavy clusters can be reasonably well described within the excluded-volume 
or Thomas-Fermi approximation \cite{hempel-typel,raduta2} and do not 
require a modification of the meson couplings. However, the presence of 
heavy clusters will have an indirect effect on the light cluster abundancies.
 In Refs. \cite{pais15,avancini17}, the
  authors have considered light clusters coexisting with a heavy
cluster and a proton-neutron background gas and showed that the
presence of heavy clusters shifts the light cluster
Mott densities to larger values. Moreover, it was also shown that above a
density $\sim 10^{-3}$ fm$^{-3}$ for $T=5$ MeV and $\sim 10^{-2}$
fm$^{-3}$ for $T=10$ MeV the presence of heavy clusters reduces the
light cluster mass fractions. Introducing heavy clusters will,
therefore, have an important effect precisely in the region where changing the
coupling $g_{si}$ has the largest effect, indicating that a more
complete study which includes heavy clusters, must be carried out. 

Notice however, that we do not have experimental results on the Mott
densities that could put constraints on the model.  On the other hand, we do have
experimental results for the chemical  equilibrium
constants (EC) obtained measuring cluster formation in heavy ion
collisions. In the next section we will compare these quantities
extracted from experimental data
with the predictions of our models. In the experimental sample, 
particles are emitted in the mid-rapidity region of a collision between relatively light ions 
under a strong radial flow, implying that no heavy clusters are present. Therefore, it makes sense that we
calculate these quantities only considering the light clusters, as previously done also in Ref. \cite{hempel15}.

\subsection{Equilibrium constants}

\begin{figure*}[!htbp]
  \begin{tabular}{cc}
	\includegraphics[width=0.75\textwidth]{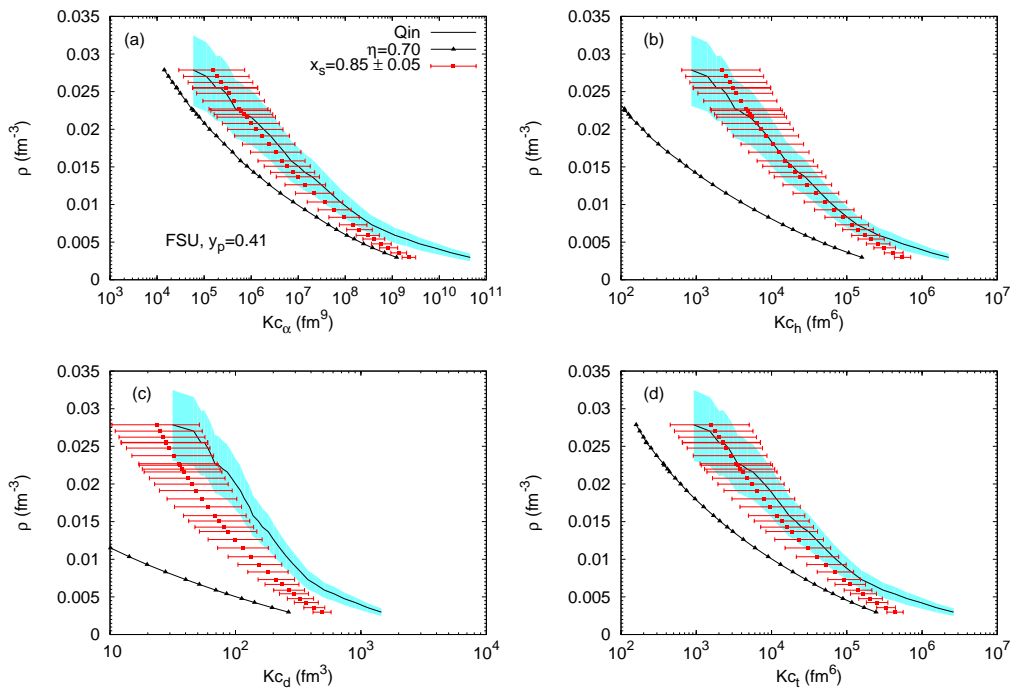}
 \end{tabular}
 \caption{(Color online) Chemical equilibrium constants of $\alpha$ (a), helion
   (b), deuteron (c), and triton (d)   for FSU, and $y_p=0.41$, and for the $\eta=0.70$
   (black squares) fitting (check Ref. \cite{avancini17} for the complete parameter sets), and the universal $g_{sj}$ fitting with $g_{sj}=(0.85\pm0.05)\, A_j \,g_s$, (red dotted lines). 
   The experimental results
   of Qin et al \cite{qin12} (light blue region) are also shown. 
} 
\label{fig7}
\end{figure*}

A very interesting constraint at high density and temperature 
was recently proposed from heavy ion collision experiments in Ref.~\cite{qin12}.

This constraint should be taken with some caution, because the systematics of such measurement are very hard to estimate. 
 First, the freeze-out concept has been used to describe the expanding fireball which is a strongly nonequilibrium process. In addition, the heavy ion reaction used involves small nuclei, and might be sensitive to important finite size and finite particle number effects. Moreover the detection was performed in a very limited angular range, and it is far from being clear that the  transient system formed during the collision and subject to a strong radial field is compatible with the laws of thermodynamical equilibrium. Finally, proton fraction ($y_p$), density and temperature are not directly observables, and a strong model dependence is associated to the determination of these variables.

Still, these data are presently the unique existing constraint on in-medium modifications of light particle yields at high temperature, and in the following  we will, therefore,
examine how well our parametrizations can reproduce the  equilibrium
constants (EC) reported in Ref.~\cite{qin12}.

With the same set of couplings determined in the last section, we
calculate the chemical equilibrium constants 
\begin{equation}
K_c[j]=\frac{\rho_j}{\rho_n^{N_j}\rho_p^{Z_j}}
\label{kc}
\end{equation}
where $\rho_j$ is the number density of cluster $j$, with neutron
number $N_j$  and  proton number $Z_j$, and $\rho_p$, $\rho_n$ are,
respectively, the number densities of free protons and neutrons.
We will calculate the EC for a proton fraction equal to 0.41, as was
assumed in \cite{qin12,hempel15}. It has, however, been shown that
dependence of the EC on the proton fraction is very small, see Ref. \cite{avancini17} and Ref. \cite{hempel15} for a discussion on this point.

 In Fig. \ref{fig7}, we show the chemical equilibrium constants for
 all the light clusters considered, taking the range of the couplings
 to be $g_{sj}=(0.85\pm0.05)\, A_j g_s$. In this Figure, we also show
 results for the parametrization  obtained in Ref. \cite{avancini17}
 for $\eta=0.70$ (black squares). This model
 describes quite well the experimental results for the
 $\alpha$ cluster, because the parametrization was fitted to the
 $\alpha$-equilibrium constants. However, it completely fails to
 reproduce the EC of the deuteron and the triton.

Taking the coupling fractions $x_{sj}=0.85\pm0.05$ essentially describes the experimental
   equilibrium constants. We have checked that $x_s=0.95$ would be  too large. 

In Ref. \cite{hempel15}, the authors have compared the EC calculated within
different models with the experimental data of  Qin {\it et
al.} \cite{qin12}, and, in particular, tested the cluster formation and the
in-medium modification of the cluster properties. They have shown that
the QS formalism gives an excellent description of the experimental
EC. Also, the generalized relativistic-density functional (gRDF)
discussed in \cite{typel10,typel12-v} gives a very good description of the EC. The
gRDF model is a meson-exchange based effective
relativistic mean-field model which includes as  degrees of freedom  nucleons, light nuclei,
and heavy nuclei, and considers
medium dependent binding energy shifts of nuclei. 
 In the gRDF model,  the in-medium binding energy
shifts were fitted to  the QS results for the light nuclei and a
Thomas-Fermi calculation for the heavy ones, and include a temperature dependence.
Another model giving
a good description of the EC is the Hempel and Schaffner-Bielich (HS) EoS \cite{hempel}
with the DD2 interaction of Typel {\it et al.} \cite{typel10}. This
model consists on a mixture of nuclei and unbound nucleons in nuclear
statistical equilibrium, the nucleons being described within a
relativistic mean-field, in this case the DD2 interaction.

Our present approach gives results similar to the last two approaches and even
to the QS prediction, except for the deuteron.  Therefore, we consider that our
proposal for the effective description of the in-medium effects on the
light clusters,  given by the temperature independent binding energy shifts
defined in Eq. (\ref{deltaB}), is justified, and presents the extra advantage to be 
applicable also to heavier hydrogen and helium isotopes which are predicted to be abundant 
in high temperature neutron rich matter \cite{raduta2}. 

Instead of the DD2 version used in other approaches, we used the FSU version of the RMF model. We suppose that the different RMF models will not show a large effect on the results in the low-density region ($n<$ 0.03 fm$^{-3}$) considered here. Larger deviations between the different versions of the RMF model are expected for nucleon densities above the saturation density. 

This experimental data seem to put extra constraints,
that together with VEoS, suggest that a good universal
coupling for all clusters is $g_{sj} = (0.85 \pm
0.05)\, A_j\, g_s$. For the deuteron, the  experimental data seem
to be described by the upper limit $x_s=0.9$. Possibly a more 
detailed approach would allow for a different coupling
$g_{sj}$ for each cluster. According to the experimental data, the deuteron seems to be  more adequately described by
taking $g_{sd} = (0.9 \pm 0.03) 2 g_s$. It should be stressed
that the deuteron is only a weakly bound state consisting  of two nucleons.

\section{Conclusions} \label{IV}

In this paper we have proposed a simple parametrization of in-medium effects acting 
on light clusters, in the framework of the relativistic mean field approximation. 
The interactions of the clusters with the surrounding medium are described with 
a phenomenological modification of the coupling constant to the $\sigma$-meson. 
A coupling proportional to the cluster size is proposed, with a  correction factor 
which is obtained imposing that the cluster fractions exhibit the correct virial behavior 
in the low density limit.
The phenomenon of cluster dissolution at high density is described introducing 
a simple binding energy shift which can be analytically derived in the Thomas-Fermi 
approximation as the energetic counterpart of the classical exclusion volume mechanism. 
With a universal cluster coupling fraction {$x_s=0.85\pm 0.05$}, 
we reproduce reasonably well both the virial limit and the equilibrium constants 
extracted from heavy ion data. A correct description of the deuteron is probably out
of scope within the mean-field approximation.
Our results are qualitatively similar to the ones 
obtained with more microscopic approaches in Refs.~\cite{typel10,Roepke15}, 
and have the advantage of being applicable also to other light clusters, 
which might have a non negligible contribution in warm asymmetric stellar matter, 
as it is produced in proto-neutron stars, supernova {environments}, and neutron star mergers.

The uncertainty in the coupling has a negligible influence for densities below 
$\approx 10^{-2}$ fm$^{-3}$. At higher densities the dispersion becomes larger 
and the predictions of the different models show a considerable
deviation. 
The dissolution density can thus vary of a factor 2-3 depending on the model, as well as on the choice 
of the coupling constants. The evolution of the dissolution density with temperature 
also varies approximately by a factor of two within the error bar of
the couplings. Besides, as discussed above, heavier clusters and the formation of pasta structures also become of relevance and should be explicitly included.

More sophisticated prescriptions allowing for different couplings
for each cluster, a non-linear mass dependence, or an explicit temperature dependence could be envisaged. 
However, to improve the present phenomenological description and fix these additional parameters, 
extra constraints from experimental data and/or microscopic calculations 
around the dissolution density will be needed.

\section*{ACKNOWLEDGMENTS}

We thank Stefan Typel for providing the Virial EoS. This work was partly supported by the FCT (Portugal) Project No. UID/FIS/04564/2016, and by former NewCompStar, COST Action MP1304. H.P. is supported by FCT (Portugal) under Project No. SFRH/BPD/95566/2013. She is very thankful to F.G. and her group at LPC (Caen) for the kind hospitality
during her stay there within a NewCompStar STSM, where this work started.

\end{document}